\documentclass{PoS}

\def\be{\begin{equation}}
\def\ee{\end{equation}}
\def\bea{\begin{eqnarray}}
\def\eea{\end{eqnarray}}
\def\beqa{\begin{eqnarray}}
\def\eeqa{\end{eqnarray}}

\def\D0bar{\overline D{}^0}
\def\DDbar{D{}^0-\overline D{}^0}

\def\beq{\begin{equation}}
\def\eeq{\end{equation}}
\def\bea{\begin{eqnarray}}
\def\eea{\end{eqnarray}}

\newcommand{\DzDzb}{D^0-\overline{D^0}}

\title{Searching for New Physics with Charm}

\ShortTitle{Searching for New Physics with Charm}

\author{\speaker{Alexey A Petrov}\\
        Department of Physics and Astronomy\\
        Wayne State University\\
        Detroit, MI 48201\\
        and \\
        Michigan Center for Theoretical Physics\\
        University of Michigan\\
        Ann Arbor, MI 48109\\
        USA\\
        E-mail: \email{apetrov@wayne.edu}}

\abstract{I provide a comprehensive review of indirect searches for New Physics with charmed mesons. I discuss current 
theoretical and experimental challenges and successes in understanding decays and mixings of those mesons.
I argue that in many New Physics scenarios strong constraints, that surpass those from other search techniques, could 
be placed on the allowed model parameter space using the existent data from studies of charm transitions. This 
has direct implications for direct searches of physics beyond the Standard Model at the LHC.}

\FullConference{12th International Conference on B-Physics at Hadron Machines - BEAUTY 2009\\
		 September 07 - 11 2009\\
		 Heidelberg, Germany}

\begin{document}

\section{Introduction}

Processes involving charm quarks provide a unique place to search for indirect effects of New Physics (NP).
They furnish a rather unique access to processes in the up-quark sector, which is not yet available in the decays of 
top quarks: neutral mesons containing charm quark are the only mesons in that sector that can have flavor oscillations.  

A distinctive feature of charmed quark systems is that they involve a "not-so-heavy" charm quark. That means that
all charmed hadrons' masses, ${\cal O}(2\mbox{~GeV})$, are placed in the middle of the region where 
non-perturbative hadronic physics is operative. While this fact does not markedly affect theoretical 
description of leptonic and semileptonic decays of charmed hadrons, it poses significant challenges in 
the analyses of their hadronic transitions. There is a great deal of optimism, however, that abundant 
experimental data would provide some hints on the structure of charm hadronic decays.
In addition, recent advances in lattice Quantum Chromodynamics (QCD) and other non-perturbative techniques 
provide us with hope that those problems will eventually be overcome. 
One can place types of searches for New Physics in the charm quark sector into three distinct categories,

\begin{enumerate}\addtolength{\itemsep}{-0.5\baselineskip}
\item 
Searches in the processes that are {\it allowed} in the Standard Model. 

In light of what was said above, it might be difficult to identify New Physics contributions to charm-initiated processes 
that are allowed in the Standard Model (SM). Yet, it is still possible. Searches of that type include testing relations 
among SM-allowed processes that are known to hold only in the SM, but not necessarily in models beyond the 
Standard Model. An usual example, which has traditionally been employed in B-physics, is testing
Cabbibo-Kobayashi-Maskawa (CKM) triangle relations. Another example is to look for processes 
where QCD seem to be under theoretical control, such as leptonic decays of $D$-mesons, $D_q \to \ell \bar \nu$. 
\item
Searches in the processes that are {\it forbidden} in the Standard Model {\it at tree level}.

Processes that involve flavor-changing neutral current (FCNC) interactions that change charm quantum number by one 
or two units do not occur in the Standard Model at tree level, as terms that mediate such interactions are absent 
from the SM Lagrangian. However, they can happen in the Standard Model at one loop level, which makes them rather rare.
Processes like that can receive New Physics contributions from both tree-level interactions mediated by new interactions, as 
well from one-loop corrections with NP particles. Processes of that types include $\DDbar$ mixing, or
inclusive and exclusive transitions mediated by $c \to u \gamma$ or  $c \to u \ell \bar \ell$. Lastly, searches for
CP-violation in charm decays could be included here as well.
\item
Searches in the processes that are {\it forbidden} in the Standard Model.

There are a set of processes that, while allowed by space-time symmetries, are forbidden in the Standard Model.
Processes of that type are so rare that searches for their signatures require incredibly high statistics experiments.
Their observation, however, would constitute a high-impact discovery, as it would unambiguously point towards
physics beyond the Standard Model. Examples include searches for lepton- and 
baryon-number-violating transitions such as $D^0 \to n \bar\nu$ or $D^0 \to \bar p e^+$, etc.
\end{enumerate}
In what follows we shall review theoretical status of searches for New Physics in charm decays.

\section{Processes allowed in the Standard Model}

\subsection{Leptonic decays of $D^+$ and $D_s$ mesons}

Due to their overall simplicity, charm leptonic decays could serve as nice laboratories to
study New Physics, as the Standard Model "background" depend on a single non-perturbative 
parameter, the decay constant $f_{D_q}$,
\beq
\langle 0 | \bar q\gamma^\mu \gamma_5 c | D_q \rangle = i f_{D_q} p_D^\mu.
\eeq
In quark model, $f_{D_q}$ parameterizes the amplitude of probability for the heavy and a
light quark to ``find each other'' in a meson. Thus, in the SM, the leptonic decay width can be 
written as
\begin{equation}\label{LeptonicDecay}
\Gamma(D_q\to \ell\nu) = {G_F^2\over
8\pi}f_{D_q}^2m_{\ell}^2M_{D_q}  \left(1-{m_{\ell}^2\over
M_{D_q}^2}\right)^2 \left|V_{cq}\right|^2~~~,
\end{equation}
where $q=d,s$ for $D^+$ or $D_s$ states respectively, $M_{D_q}$ is
the $D_q$ mass, $m_{\ell}$ is the mass of the final state lepton,
and $|V_{cq}|$ is the $CKM$ matrix element associated with the $c
\to q$ transition. Due to helicity suppression the rate goes as
$m_\ell^2$, which plays a role in NP searches as many NP models could have a different 
parametric dependence on $m_\ell^2$ (or not at all).
Thus, provided an accurate calculation of the SM contribution (and, in 
particular, $f_{D_q}$) is available, one can place rather tight constraints 
on some models of New Physics. 
\begin{center}
\begin{table}[hbtp]
\begin{center}
\begin{tabular}{llcc}
\hline \hline 
Experiment & Mode & ${\cal B}(x10^3)$ & $f_{D_s}$\ (MeV)\\
\hline 
CLEO-c & $\mu ^+\nu_{\mu}$  & $5.94\pm 0.66\pm 0.31$  & $264\pm 15 \pm 7$ \\
CLEO-c & $\tau ^+\nu_{\tau}$  & $80.0\pm 13.0\pm 4.0$   & $310\pm 25\pm 8$\\ 
CLEO-c & $\tau ^+\nu_{\tau}$  & $61.7 \pm 7.1 \pm 3.6 $    &  $275 \pm 10 \pm 5$ \\
CLEO-c & combined & ~~ & $274 \pm 10 \pm 5$\\
Belle & $\mu ^+\nu_{\mu}$ &  $6.44\pm 0.76\pm 0.52 $ & $279\pm16 \pm 12$\\
Average & ~~ & ~~& $275\pm 10$ \\
\hline\hline
Theory & ~~~ & ~~~ & $f_{D_s}$\ (MeV)\\
\hline
HPQCD & ~~~ & ~~~ & $241 \pm 3$ \\
FNAL &  ~~~ & ~~~ & $249 \pm 3 \pm 16$ \\
\hline
\end{tabular}
\end{center}
\caption{Experimental/theoretical results for $D_s$ decay constant before 
2009 (see \cite{Artuso:2008vf} for more details).}
 \label{TableLeptonicBefore2009}
\end{table}
\end{center}
Accurate calculations of non-perturbative QCD parameters are very challenging, for which 
lattice QCD  represents an appealing approach. In the past a big stumbling block in the lattice studies 
of QCD has been the inclusion of dynamical quark effects, i.e. "unquenching" lattice QCD. In the recent years, 
technical developments such as highly improved actions of QCD and the availability of 
``2+1flavor'' MILC configurations with 3 flavors of improved staggered quarks have lead to results with 
much higher accuracy and allowed for consistent estimate of both statistical and systematic errors
involved in the simulations. Two groups have reported  charm decay constant calculations with three 
dynamical quark flavors, the Fermilab/MILC Lattice collaboration~\cite{Aubin:2005ar} and the
HPQCD collaboration~\cite{Follana:2007uv}. Their results, along with experimental measurements from 
CLEO-c and Belle, are presented in Table \ref{TableLeptonicBefore2009}.
As can be easily seen, there is a 3.6$\sigma$ discrepancy between HPQCD-predicted and 
experimentally extracted values of $f_{D_s}$, which could in principle be due to New Physics interactions. 
This is because $f_{D_s}$ was extracted from experimental data assuming only SM interactions.
Note that theoretical predictions and experimental extractions for $f_{D^+}$ are consistent 
with each other, the discrepancy is only observed in the $D_s$ system.

The possibility of New Physics being responsible for this discrepancy has been 
studied in~\cite{Dobrescu:2008er} and subsequently by many authors (see~\cite{Kronfeld:2009cf} for a 
recent summary). In principle, leptonic decays could be sensitive probes of NP interactions mediated by
charged particles. Models with an extended Higgs sector, which include new charged scalar states, 
or models with broken left-right symmetry, which include massive vector $W^\pm_R$ states, are
examples of such interactions. To account for New Physics, one can make a 
substitution~\cite{Kronfeld:2009cf} 
\beq
G_F V_{cs}^* m_\ell \to
G_F V_{cs}^* m_\ell + G_A^\ell m_\ell + G_P^\ell \frac{m_{D_s}^2}{m_c + m_s}
\eeq
in Eq. ~(\ref{LeptonicDecay}) for the $D_s$. Here  $G_A^\ell$ and $G_P^\ell$ parameterize
new couplings and masses of NP interactions.

Indeed, NP contribution to the $c \to q \ell \nu$ interaction would affect other processes, such as 
leptonic $D^+ \to \ell\nu$ and semileptonic $D \to M \ell\bar\nu$ decays. It is quite hard to satisfy all
constraints from those processes simultaneously~\cite{Kronfeld:2009cf} in many popular models 
of New Physics. Besides, new experimental results from CLEOc~\cite{Alexander:2009ux} lead 
to a new experimental average reported by Heavy Flavor Averaging Group (HFAG)~\cite{HFAG},
\beq
f_{D_s}= 256.9 \pm 6.8 ~\mbox{MeV},
\eeq
and new lattice QCD predictions (for various numbers of sea-quark flavors $n_f$) reported at 
the Lattice-2009 conference by Fermilab/MILC collaboration and by European Twisted-Mass 
Collaboration (ETMC)~\cite{Lattice09} 
\begin{eqnarray}
f_{D_s} &=& 260 \pm 10 ~\mbox{MeV}~~[n_f=2+1]~~ \mbox{(FNAL/MILC)}, 
\nonumber \\
f_{D_s} &=& 244 \pm 8 ~\mbox{MeV}~~~~[n_f=2]~~~~~~~~ \mbox{(ETMC)}
\end{eqnarray}
cast a serious doubt that this discrepancy is caused by New Physics.

There are excellent prospects for further insights into the "$f_{D_s}$-problem." Besides new lattice 
evaluations of this quantity by the same and other collaborations (for instance, with possible improvements
on new MILC ensembles with $n_f=2+1+1$, i.e. including charm sea quarks), new measurements
with a percent accuracy will be available from BES-III collaboration in a few years~\cite{Asner:2008nq}. 
This, together with continuous improvement of BaBar and Belle results, should provide a 
resolution of the "$f_{D_s}$-problem."

\subsection{CKM triangle relations in charm}

Another way to search for New Physics in the SM-allowed processes is to test relations that
only hold in the SM, but not necessarily in general. An example of such relation is a
CKM "charm unitarity triangle" relation. 
\beq\label{CharmTriangle}
V_{ud}^* V_{cd} + V_{us}^* V_{cs} + V_{ub}^* V_{cb} = 0 
\eeq
Relations like Eq.~(\ref{CharmTriangle}) hold in the SM due to a single phase of the CKM matrix 
driving CP-violation in the SM, which is not always so in general BSM models. Moreover, 
processes that are used to extract CKM parameters in Eq.~(\ref{CharmTriangle}) can 
be affected by New Physics, which might lead to difference in the shape of the triangle extracted 
from different transitions.

In fact, there are several unitarity triangles that involve charm inputs~\cite{Bigi:1999hr}.
Since all CP-violating effects in the flavor sector of the SM are related to the single phase 
of the CKM matrix, all of the CKM unitarity triangles, including the one in Eq.~(\ref{CharmTriangle}), 
have the same area, $A=J/2$, where $J$ is the Jarlskog invariant. This fact could 
provide a non-trivial check of the Standard Model, if measurements of all sides of these triangles 
are performed with sufficient accuracy and then compared to areas of other CKM unitarity triangles.

Unfortunately, the ``charm triangle'' is rather ``squashed'', with one side being 
much shorter then the other two. In terms of the Wolfenstein parameter 
$\lambda=0.22$, the relation in Eq.~(\ref{CharmTriangle}) has one side ${\cal O}(\lambda^5)$ with 
the other two being ${\cal O}(\lambda)$. This triangle relation is however quite interesting because 
all measurements needed to extract the CKM matrix elements in Eq.~(\ref{CharmTriangle}) come from 
the tree-level processes. Thus, its area should be a measure of CP-violation
in the SM, which can be compared to the area of the more familiar "B-physics triangle",
\beq\label{BTriangle}
V_{ud}^* V_{ub} + V_{cd}^* V_{cb} + V_{td}^* V_{tb} = 0 
\eeq
which receives input from loop-dominated processes like B-mixing and 
whose area squared is $A_c^2=(2.32 \pm 0.31)\times 10^{-10}$. Compared to 
this, the area of the  "charm unitarity triangle" in Eq.~(\ref{CharmTriangle}) is 
$A_c^2 = (-1.34\pm 5.46)\times 10^{-6}$ (obtained using inputs from~\cite{PDG}), which is 
clearly not precise enough for meaningful comparison.

In addition, relations like $\left|V_{cd}\right|^2+\left|V_{cs}\right|^2+\left|V_{cb}\right|^2=1$
could be tested. It could provide an interesting cross-check on the value of $V_{cb}$ 
extracted in B-decays, if sufficient accuracy on the experimental measurement of
$V_{cd}$ and $V_{cs}$ is achieved. It is however unlikely that the required accuracy 
will be achieved in the near future. 

\section{Processes forbidden in the Standard Model at tree level}

Processes forbidden in the SM at tree level involve FCNC, which can manifest
themselves in rare decays and meson-anti-meson mixing. 
The phenomenon of meson-anti-meson mixing occurs in the presence of operators that
change quark flavor by two units~\cite{Artuso:2008vf}. While those operators can be generated 
in the Standard Model at one loop, they can also be generated in its many possible extensions. 
With the potential window to discern large NP effects in the charm sector as well as
the anticipated improved accuracy for future mixing measurements, the motivation for a 
comprehensive up-to-date theoretical analysis of New Physics contributions to $D$ meson 
mixing is compelling.

\subsection{New Physics in $\DzDzb$ mixing}

The presence of $\Delta C = 2$ operators produce off-diagonal terms in the meson-anti-meson mass 
matrix, so that the basis of flavor eigenstates no longer coincides with the basis of mass eigenstates. 
Those two bases, however, are related by a linear transformation,
\beq
|D_{1\atop 2} \rangle = p | D^0 \rangle \pm q | \overline{D}^0 \rangle,
\eeq
where the complex parameters $p$ and $q$ are obtained from diagonalizing the 
$\DDbar$ mass matrix. Neglecting CP-violation leads to $p=q=1/\sqrt{2}$.
The mass and width splittings between mass eigenstates are 
\beq\label{XandY}
x_D= \frac{m_1-m_2}{\Gamma_D}, \qquad y_D=\frac{\Gamma_1-\Gamma_2}{2 \Gamma_D},
\eeq
where $\Gamma_{\rm D}$ is the average width of the two neutral $D$ meson mass eigenstates.  
Because of the absence of superheavy down-type quarks destroying Glashow-Iliopoulos-Maiani 
(GIM) cancellation, it is expected that $x_D$ and $y_D$ should be rather small in the Standard Model.
The quantities which are actually measured in experimental determinations of the mass and width differences, 
are $y_{\rm D}^{\rm (CP)}$ (measured in time-dependent $D \to KK, \pi\pi$ analyses), 
$x_{\rm D}'$, and $y_{\rm D}'$ (measured in $D \to K\pi$ and similar transitions), are defined as
\bea
y_{\rm D}^{\rm (CP)} &=&  y_{\rm D} \cos\phi - x_{\rm D}\sin\phi
\left(\frac{A_m}{2}-A_{prod}\right) \ \ , \nonumber \\
x_D' &=& x_D\cos\delta_{K\pi} + y_D\sin\delta_{K\pi} \ \ ,
\\
y_D' &=& y_{\rm D} \cos \delta_{K\pi} - x_{\rm D}
\sin\delta_{K\pi} \ \ ,
\nonumber
\label{y-defs}
\eea
where
$A_{prod} = \left(N_{D^0} - N_{{\overline D}^0}\right)/
\left(N_{D^0} + N_{{\overline D}^0}\right)$ is the so-called
production asymmetry of $D^0$ and $\overline{D}^0$ (giving
the relative weight of $D^0$ and ${\overline D}^0$ in the
sample) and $\delta_{K\pi}$ is the strong phase difference between
the Cabibbo favored and double Cabibbo suppressed
amplitudes~\cite{Bergmann:2000id}, which  can be measured in 
$D\to K\pi$ transitions. A fit to the current database of experimental analyses 
by the Heavy Flavor Averaging Group (HFAG) gives~\cite{ExperimentalAnalyses,HFAG}
\beqa\label{hfag}
& & x_{\rm D} = 0.0100^{+0.0024}_{-0.0026}~, ~~\qquad 
y_{\rm D} = 0.0076^{+0.0017}_{-0.0018} \  \ ,
\nonumber \\
& & 1 - |q/p| = 0.06 \pm 0.14, \quad 
\phi = -0.05 \pm 0.09,
\eeqa
where $\phi$ is a CP-violating phase. It is important to note that the size of the signal allows to 
conclude that the former "smoking gun" signal for New Physics in $\DDbar$ mixing, 
$x \gg y$ no longer applies. Also, CP-violating is charm is clearly small. 
The question that arises now is how to use available data to probe for physics beyond the Standard Model.

Theoretical predictions for $x_D$ and $y_D$ obtained in the framework of the Standard Model 
are quite complicated. I will not be discussing those here, instead referring the interested reader to recent 
reviews~\cite{Artuso:2008vf}. It might be advantageous to 
note that there are two approaches to describe $\DDbar$ mixing, neither of which give very 
reliable results because $m_c$ is in some sense intermediate between heavy and light. 

Let me introduce a scale $\Lambda \sim 1$~GeV to be a scale characteristic of the strong interactions.
The "inclusive" approach~\cite{Petrov:1997ch,Inclusive} is based on the operator product 
expansion (OPE) in the formal limit $m_c \gg \Lambda$, where $x_D$ and $y_D$ can be expanded in 
terms of matrix elements of local operators. The use of the OPE relies on local quark-hadron duality, and on 
$\Lambda/m_c$ being small enough to allow a truncation of the series after the first few terms. 
This, however, is not realized in $\DDbar$ mixing, as the leading term in $1/m_c$ is suppressed by four and 
six powers of the strange quark mass for $x_D$ and $y_D$ respectively. The parametrically-suppressed 
higher order terms in $1/m_c$ can have less powers of $m_s$, thus being more important 
numerically~\cite{Inclusive}. This results in reshuffling of the OPE series, making it a triple expansion in
$1/m_c$, $m_s$, and $\alpha_s$. The (numerically) leading term contains over twenty matrix 
elements of dimension-12, eight-quark operators, which are difficult to compute reliably. A naive power 
counting then yields $x_D, y_D < 10^{-3}$. The "exclusive" approach~\cite{Exclusive} 
more realistically assumes $m_c \simeq \Lambda$ and sums over intermediate hadronic states. Since there 
are cancellations between states within a given $SU(3)$ multiplet, one needs to know the contribution of 
each state with high precision. However, $D$ meson is not light enough to have only a few open decay
channels. In the absence of sufficiently precise data one is forced to use some assumptions. Large 
effects in $y_D$ appear for decays close to $D$ threshold, where an analytic expansion in $SU(3)_F$ 
violation is no longer possible. Thus, even though theoretical calculations of $x_D$ and $y_D$ are quite 
uncertain, the values $x_D \sim y_D \sim 1\%$ are natural in the Standard Model~\cite{Falk:2001hx}. 

It then appears that experimental results of Eq.~(\ref{hfag}) are consistent with the SM predictions. Yet, 
those predictions are quite uncertain to be subtracted from the experimental data to precisely
constrain possible NP contributions. In this situation the following approach can be taken. One can neglect the 
SM contribution altogether and assume that NP saturates the experimental result. 
This way, however, only an upper bound on the NP parameters can be placed. A subtlety of this method is related to 
the fact that the SM and NP contributions can have either the same or opposite signs. While the sign of the SM 
contribution cannot be calculated reliably due to hadronic uncertainties, $x_D$ computed within a given 
NP model can be determined. This  stems from the fact that NP contributions are generated by 
heavy degrees of freedom making short-distance OPE reliable. This means that only the part of parameter 
space of NP models that generate $x_D$ of the same sign as observed experimentally can be reliably 
constrained.

Any NP degree of freedom will generally be associated with a generic heavy mass scale $M$, 
at which the NP interaction will be most naturally described.  At the scale $m_c$ of the charm mass, 
this description will have been modified by the effects of QCD, which should be taken into account.  
In order to see how NP might affect the mixing amplitude, it is instructive to consider off-diagonal 
terms in the neutral D mass matrix,
\begin{eqnarray}\label{M12}
 \left (M - \frac{i}{2}\, \Gamma\right)_{12} =
 \frac{1}{2 M_{\rm D}} \langle \D0bar | 
{\cal H}_w^{\Delta C=-2} | D^0 \rangle 
+  \frac{1}{2 M_{\rm D}} \sum_n {\langle \D0bar | {\cal H}_w^{\Delta
  C=-1} | n \rangle\, \langle n | {\cal H}_w^{\Delta C=-1} 
| D^0 \rangle \over M_{\rm D}-E_n+i\epsilon}\,
\end{eqnarray}
where the first term contains ${\cal H}_w^{\Delta C=-2}$, which is an 
effective $|\Delta C| = 2$ hamiltonian, represented by a set of operators that are local at 
the $\mu \simeq m_D$ scale. Note that a $b$-quark also gives a (negligible) contribution
to this term. This term only affects $x_D$, but not $y_D$. 

The second term in Eq.~(\ref{M12}) is given by a double insertion of the effective $|\Delta C| = 1$ 
Hamiltonian ${\cal H}_w^{\Delta C=-1}$. This term is believed to give dominant contribution to 
$\DDbar$ mixing in the Standard Model, affecting both $x$ and $y$. It is also generally believed that
NP cannot give any sizable contribution to this term, since ${\cal H}_w^{\Delta C=-1}$ Hamiltonian 
also mediates non-leptonic $D$-decays, which should then also be affected by this NP contribution.
To see why this is not so, consider a non-leptonic $D^0$ decay amplitude, $A[D^0 \to n]$, which includes 
a small NP contribution, $A[D^0 \to n]=A_n^{\rm (SM)} + A_n^{\rm (NP)}$. Here, $A_n^{\rm (NP)}$ is assumed 
to be smaller than the current experimental uncertainties on those decay rates. This ensures that NP 
effects cannot be seen in the current experimental analyses of non-leptonic D-decays. Then, $y_{\rm D}$ is
\begin{eqnarray}\label{schematic}
y_{\rm D} &\simeq& \sum_n \frac{\rho_n}{\Gamma_{\rm D}} 
A_n^{\rm (SM)} A_n^{\rm (SM)}
+ 2\sum_n \frac{\rho_n}{\Gamma_{\rm D}}
A_n^{\rm (NP)} A_n^{\rm (SM)} \ \ . 
\end{eqnarray}
The first term of Eq.~({schematic}) represents the SM contribution to $y_{\rm D}$. The SM contribution to $y_{\rm D}$ is 
known to vanish in the limit of exact flavor $SU(3)$. Moreover, the first order correction is also absent, 
so the SM contribution arises only as a {\it second} order effect~\cite{Falk:2001hx}. This means that
in the flavor $SU(3)$ limit the lifetime difference $y_{\rm D}$ is dominated by the second term in 
Eq.~(\ref{schematic}), i.e. New Physics contributions, even if their contibutions are tiny in the individual decay 
amplitudes~\cite{Golowich:2006gq}!  A realistic calculation reveals that NP contribution to $y_{\rm D}$ can be as large 
as several percent in R-parity-violating SUSY models~\cite{Petrov:2007gp} or as small as $\sim 10^{-10}$ 
in the models with interactions mediated by charged Higgs particles~\cite{Golowich:2006gq}.

As mentioned above, heavy BSM degrees of freedom cannot be produced in charm meson decays, but 
can nevertheless affect effective $|\Delta C| = 2$ Hamiltonian by changing Wilson coefficients and/or introducing 
new operator structures. By integrating out those new degrees of freedom associated with new interactions at a 
high scale $M$, we are left with an effective hamiltonian written in the form of a series of operators of increasing 
dimension. It turns out that a model-independent study of NP $|\Delta C| = 2$ contributions is possible, as 
any NP model will only modify Wilson coefficients of those operators~\cite{Golowich:2007ka,Gedalia:2009kh},
\beq\label{SeriesOfOperators}
{\cal H}_{NP}^{|\Delta C| = 2}  =\frac{1}{M^2} \left[ 
\sum_{i=1}^8  {\rm C}_i (\mu) ~ Q_i  \right],
\eeq
where ${\rm C}_i$ are dimensionless Wilson coefficients, and the $Q_i$ are the effective operators:
\beqa
\begin{array}{l}
Q_1 = (\overline{u}_L^\alpha \gamma_\mu c_L^\alpha) \ 
(\overline{u}_L^\beta \gamma^\mu c_L^\beta)\ , \\
Q_2 = (\overline{u}_R^\alpha c_L^\alpha) \ 
(\overline{u}_R^\beta c_L^\beta)\ , \\
Q_3 = (\overline{u}_R^\alpha c_L^\beta) \ 
(\overline{u}_R^\beta c_L^\alpha) \ , \\
Q_4 = (\overline{u}_R^\alpha c_L^\alpha) \ 
(\overline{u}_L^\beta c_R^\beta) \ ,
\end{array}
\qquad
\begin{array}{l}
Q_5 = (\overline{u}_R^\alpha c_L^\beta) \ 
(\overline{u}_L^\beta c_R^\alpha) \ , \\
Q_6 = (\overline{u}_R^\alpha \gamma_\mu c_R^\alpha) \ 
(\overline{u}_R^\beta \gamma^\mu c_R^\beta)\ , \\
Q_7 = (\overline{u}_L^\alpha c_R^\alpha) \ 
(\overline{u}_L^\beta c_R^\beta)\ , \\
Q_8 = (\overline{u}_L^\alpha c_R^\beta) \ 
(\overline{u}_L^\beta c_R^\alpha) \ \ ,
\end{array}
\label{SetOfOperators}
\eeqa
where $\alpha$ and $\beta$ are color indices. In total, there are eight possible operator structures that exhaust the
list of possible independent contributions to $|\Delta C|=2$ transitions.
Note that earlier Ref.~\cite{Golowich:2007ka} used a slightly different set of operators than~\cite{Gedalia:2009kh}, 
which can be related to each other by a linear transformation. Taking operator mixing into account, 
a set of constraints on the Wilson coefficients of Eq.~(\ref{SeriesOfOperators}) can be placed,
\beqa
\begin{array}{l}
\left| C_1 \right| \leq 5.7 \times 10^{-7} \left[\frac{M}{1~\mbox{TeV}} \right]^2, 
\\
\left| C_2 \right| \leq 1.6 \times 10^{-7} \left[\frac{M}{1~\mbox{TeV}} \right]^2, 
\\
\left| C_3 \right| \leq 5.8 \times 10^{-7} \left[\frac{M}{1~\mbox{TeV}} \right]^2, 
\end{array}
\qquad
\begin{array}{l}
\left| C_4 \right| \leq 5.6 \times 10^{-8} \left[\frac{M}{1~\mbox{TeV}} \right]^2, 
\\
\left| C_5 \right| \leq 1.6 \times 10^{-7} \left[\frac{M}{1~\mbox{TeV}} \right]^2.
\end{array}
\label{ConstraintsOnCoefficients}
\eeqa
The constraints on $C_6-C_8$ are identical to those on $C_1-C_3$~\cite{Gedalia:2009kh}.
Note that Eq.~(\ref{ConstraintsOnCoefficients}) implies that New Physics particles, for some
unknown reason, has highly suppressed couplings to charmed quarks. Alternatively, 
the tight constraints of Eq.~(\ref{ConstraintsOnCoefficients}) probes NP at the very high scales:
$M \ge (4-10) \times 10^3$~TeV for tree-level NP-mediated charm mixing and 
$M \ge (1-3) \times 10^2$~TeV for loop-dominated mixing via New Physics particles.

A contribution to $\DDbar$ mixing from a particular NP model can be obtained by calculating matching 
conditions for the Wilson coefficients $C_i$ at the scale $M$, running their values down to $\mu$ and
computing the relevant matrix elements of four-quark operators. This program has been executed 
in Ref.~\cite{Golowich:2007ka} for 21 well-motivated NP models, which will be actively studied at LHC. 
The results are presented in Table~\ref{tab:bigtableofresults}.
As can be seen, out of 21 models considered, only four received no useful constraints
from $\DDbar$ mixing. More informative exclusion plots can be found in that 
paper~\cite{Golowich:2007ka} as well. It is interesting to note that some models 
{\it require} large signals in the charm system if mixing and FCNCs in the strange 
and beauty systems are to be small (as in, for example, the SUSY alignment 
model~\cite{Nir:1993mx,Ciuchini:2007cw,Altmannshofer:2010ad}). 
\begin{table}[t]
\begin{center}
\begin{tabular}{|c||c|}
\hline 
Model & Approximate Constraint 
\\ \hline\hline
Fourth Generation  &\ \ $|V_{ub'} V_{cb'}|\cdot m_{b'}  
<   0.5 $~(GeV)  
\ \ \\
$Q=-1/3$ Singlet Quark  
&  $s_2\cdot m_S  < 0.27$~(GeV) 
\\
$Q=+2/3$ Singlet Quark 
&  $|\lambda_{uc}| < 2.4 \cdot 10^{-4}$ \\
Little Higgs  &  Tree: See entry for $Q=-1/3$ Singlet Quark \\
& Box: Parameter space can reach 
observed $x_{\rm D}$
\\
Generic $Z'$
&  $M_{Z'}/C > 2.2\cdot 10^3$~TeV  \\
Family Symmetries  & $m_1/f>1.2\cdot 10^{3}$~TeV 
 (with $m_1/ m_2 = 0.5$)   \\
Left-Right Symmetric   & No constraint   \\
Alternate Left-Right Symmetric  &
$M_R>1.2$~TeV ($m_{D_1}=0.5$~TeV)   \\
 & ($\Delta m/m_{D_1})/M_R>0.4$~TeV$^{-1}$\\
Vector Leptoquark Bosons 
& $M_{VLQ} > 55 (\lambda_{PP}/0.1) $~TeV  \\
Flavor Conserving Two-Higgs-Doublet 
&   No constraint \\
Flavor Changing Neutral Higgs 
&  $m_H/C>2.4\cdot 10^3$~TeV \\
FC Neutral Higgs (Cheng-Sher)  & 
$m_H/|\Delta_{uc}|>600$~GeV\\
Scalar Leptoquark Bosons  & See entry for RPV SUSY \\
Higgsless   & $M > 100$~TeV  \\
Universal Extra Dimensions & No constraint \\
Split Fermion  
& $M / |\Delta y| > (6 \cdot 10^2~{\rm GeV})$
 \\
Warped Geometries &  $M_1 > 3.5$~TeV \\
MSSM  & 
$|(\delta^u_{12})_{\rm LR,RL}| 
< 3.5 \cdot 10^{-2}$ for ${\tilde m}\sim 1$~TeV  \\
  & 
$|(\delta^u_{12})_{\rm LL,RR}| < .25 $ for ${\tilde m}\sim 1$~TeV  
\\
SUSY Alignment & ${\tilde m} > 2$~TeV  \\
Supersymmetry with RPV & 
$\lambda'_{12k} \lambda'_{11k}/m_{\tilde d_{R,k}} < 
1.8 \cdot 10^{-3}/100$~GeV
\\
Split Supersymmetry & No constraint  \\
\hline\hline
\end{tabular}
\end{center}
\vskip .05in\noindent
\caption{Approximate constraints on NP models from $D^0$ mixing (from~\cite{Golowich:2007ka}).}
\label{tab:bigtableofresults}
\end{table}
%

\subsection{New Physics in rare decays of charmed mesons}

I will call {\it rare} those decays of $D$ mesons that are mediated by quark-level FCNC transitions
$c \to u \gamma$ (rare radiative) and $c \to u \ell \ell$ (rare leptonic and semileptonic). These decays 
only proceed at one loop in the SM, so just like in $\DDbar$ mixing GIM mechanism is very effective.
Here I will concentrate on the simplest rare leptonic decays $D^0 \to \ell^+ \ell^-$. These transitions
have a very small SM contribution, so they could be very cleans probes of NP amplitudes. Other transitions
rare decays (such as $D \to \rho \gamma$, etc.) could receive rather significant SM contributions, which are
quite difficult to compute. For more information on those decays please see Refs.~\cite{Burdman:2001tf}.

Experimentally, at present, there are only the upper 
limits~\cite{PDG,Aubert:2004bs,Abt:2004hn,Acosta:2003ag} on $D^0 \to \ell^+ \ell^-$ decays,
\beqa
{\cal B}_{D^0 \to \mu^+\mu^-} \le  1.3\times 10^{-6}, \quad 
{\cal B}_{D^0 \to e^+ e^-} \le 1.2\times 10^{-6}, ~~ {\rm and}~~ 
{\cal B}_{D^0 \to \mu^\pm e^\mp} \le  8.1\times 10^{-7}.
\label{brs}
\eeqa
Theoretically, just like in the case of mixing discussed above, all possible NP contributions to $c \to u \ell^+ \ell^-$ 
can also be summarized in an effective hamiltonian,  
\beq\label{SeriesOfOperators2}
{\cal H}_{NP}^{rare}  = 
\sum_{i=1}^{10}  {\rm \widetilde C}_i (\mu) ~ \widetilde Q_i,
\eeq
where ${\rm \widetilde C}_i$ are again Wilson coefficients, and the $ \widetilde Q_i$ are the effective operators.
In this case, however, there are ten of them, 
\beqa
\begin{array}{l}
\widetilde Q_1 = (\overline{\ell}_L \gamma_\mu \ell_L) \ 
(\overline{u}_L \gamma^\mu
c_L)\ , \\
\widetilde Q_2 = (\overline{\ell}_L \gamma_\mu \ell_L) \ 
(\overline{u}_R \gamma^\mu
c_R)\ , \\ 
\widetilde Q_3 = (\overline{\ell}_L \ell_R) \ (\overline{u}_R c_L) \ , 
\end{array}
\qquad 
\begin{array}{l}
\widetilde Q_4 = (\overline{\ell}_R \ell_L) \ 
(\overline{u}_R c_L) \ , \\
\widetilde Q_5 = (\overline{\ell}_R \sigma_{\mu\nu} \ell_L) \ 
( \overline{u}_R \sigma^{\mu\nu} c_L)\ ,\\
\phantom{xxxxx} 
\end{array}
\label{SetOfOperatorsLL}
\eeqa
with five additional operators $\widetilde Q_6, \dots, \widetilde Q_{10}$ 
that can be obtained from operators in Eq.~(\ref{SetOfOperatorsLL}) by 
the substitutions $L \to R$ and $R \to L$.
It is worth noting that only eight operators contribute to
$D^0\to \ell^+\ell^-$, as 
$\langle \ell^+ \ell^- | \widetilde Q_5 | D^0 \rangle =
\langle \ell^+ \ell^- | \widetilde Q_{10} | D^0 \rangle = 0$. 
The most general $D^0 \to \ell^+ \ell^-$ decay amplitude can be written as
\beq\label{decayampl}
{\cal M} = {\bar u}({\bf p}_-, s_-) \left[ A + B \gamma_5 
\right] v({\bf p}_+, s_+) \ \ , 
\eeq
which result in the branching fractions 
\begin{eqnarray}\label{Dllgen}
& & {\cal B}_{D^0 \to \ell^+\ell^-} = 
\frac{M_D}{8 \pi \Gamma_{\rm D}} \sqrt{1-\frac{4 m_\ell^2}{M_D^2}}
\left[ \left(1-\frac{4 m_\ell^2}{M_D^2}\right)\left|A\right|^2  +
\left|B\right|^2 \right] \ \ , 
\nonumber \\
& & {\cal B}_{D^0 \to \mu^+e^-} = 
\frac{M_D}{8 \pi \Gamma_{\rm D}} 
\left( 1-\frac{ m_\mu^2}{M_D^2} \right)^2 
\left[ \left|A\right|^2  + \left|B\right|^2 \right] \ \ .
\end{eqnarray}
I neglected the electron mass in the latter expression. Any NP contribution described by 
the operators of Eq.~(\ref{SetOfOperatorsLL}) gives for 
the amplitudes $A$ and $B$, 
\begin{eqnarray}
\left| A\right|  &=& G \frac{f_D M_D^2}{4 m_c} \left[\widetilde C_{3-8} + 
\widetilde C_{4-9}\right]\ , \nonumber 
 \\
\left| B\right|  &=& G \frac{f_D}{4} \left[
2 m_\ell \left(\widetilde C_{1-2} + \widetilde C_{6-7}\right)
+  \frac{M_D^2}{m_c}
\left(\widetilde C_{4-3} + \widetilde C_{9-8}\right)
\right]\ ,  \label{DlCoeff}
\end{eqnarray}
with $\widetilde C_{i-k} \equiv \widetilde C_i-\widetilde C_k$. Any NP 
model that contribute to $D^0 \to \ell^+ \ell^-$ can be constrained
from the constraints on the Wilson coefficients in Eq.~(\ref{DlCoeff}).
\begin{center}
\begin{table}[th]
\begin{center}
\begin{tabular}{|c|c|}
\hline 
Model & ${\cal B}(D^0 \to \mu^+\mu^-)$ \\ 
\hline
Experiment & $\le 1.3 \times 10^{-6}$ \\
Standard Model (LD) & $\sim {\rm several} \times 10^{-13}$ \\
$Q=+2/3$ Vectorlike Singlet & $4.3 \times 10^{-11}$ \\
$Q=-1/3$ Vectorlike Singlet & $1 \times10^{-11}~(m_S/500~{\rm GeV})^2$ \\
$Q=-1/3$ Fourth Family & $1 \times 10^{-11}~(m_S/500~{\rm GeV})^2$ \\
$Z'$ Standard Model (LD) & $2.4 \times 10^{-12}/(M_{Z'}{\rm (TeV)})^2$ \\
Family Symmetry & $0.7 \times 10^{-18}$  (Case A)  \\
RPV-SUSY & $~4.8 \times 10^{-9}~(300~{\rm GeV}/m_{\tilde{d}_k})^2$ \\
\hline
\end{tabular}
\end{center}
\vskip .05in\noindent
\caption{Predictions for $D^0 \to \mu^+\mu^-$ branching fraction for $x_D \sim 1\%$ (from~\cite{Golowich:2009ii})}
\label{tab:corr}
\end{table}
\end{center}
It is, however, possible to go further. In particular, it might be advantageous to 
study {\it correlations} of New Physics contributions to various processes, for instance
$\DDbar$ mixing and rare decays~\cite{Golowich:2009ii}. In general, one cannot predict 
the rare decay rate by knowing just the mixing rate, even if
both $x_D$ and ${\cal B}_{D^0 \to \ell^+\ell^-}$ are dominated by a 
given NP contribution. It is, however, possible for a restricted subset of NP models~\cite{Golowich:2009ii}.
The results are presented in Table~\ref{tab:corr}.
Note that similar correlated studies can be done with other systems, for instance correlating results in $K$, $B$ and
$D$ mixing~\cite{Blum:2009sk}.

\section{"Smoking gun" signals: CP-violation in charm}

Another possible manifestation of new physics interactions in the charm
system is associated with the observation of (large) CP-violation~\cite{Artuso:2008vf,Bigi:2009jj}. 
This is due to the fact that all quarks that build up the hadronic states in weak 
decays of charm mesons belong to the first two generations. Since $2\times2$ 
Cabbibo quark mixing matrix is real, no CP-violation is possible in the
dominant tree-level diagrams which describe the decay amplitudes. 
CP-violating amplitudes can be introduced in the Standard Model by including 
penguin or box operators induced by virtual $b$-quarks. However, their 
contributions are strongly suppressed by the small combination of 
CKM matrix elements $V_{cb}V^*_{ub}$. It is thus widely believed that the 
observation of (large) CP violation in charm decays or mixing would be an 
unambiguous sign for New Physics. The SM "background" here is quite small,
giving CP-violating asymmetries of the order of $10^{-3}$.

No CP-violation has been observed in charm transitions yet. However, available 
experimental constraints of Eq.~(\ref{hfag}) can provide some tests of CP-violating NP models. 
For example, a set of constraints on the imaginary parts of Wilson coefficients of 
Eq.~(\ref{SeriesOfOperators}) can be placed,
\beqa
\begin{array}{l}
\mbox{Im} \left[C_1\right] \leq 1.1 \times 10^{-7} \left[\frac{M}{1~\mbox{TeV}} \right]^2, 
\\
\mbox{Im} \left[C_2\right] \leq 2.9 \times 10^{-8} \left[\frac{M}{1~\mbox{TeV}} \right]^2, 
\\
\mbox{Im} \left[C_3\right]  \leq 1.1 \times 10^{-7} \left[\frac{M}{1~\mbox{TeV}} \right]^2, 
\end{array}
\qquad
\begin{array}{l}
\mbox{Im} \left[C_4\right]  \leq 1.1 \times 10^{-8} \left[\frac{M}{1~\mbox{TeV}} \right]^2, 
\\
\mbox{Im} \left[C_5\right]  \leq 3.0 \times 10^{-8} \left[\frac{M}{1~\mbox{TeV}} \right]^2.
\end{array}
\eeqa
Just like the constraints of Eq.~(\ref{ConstraintsOnCoefficients}), they give a sense of
how NP particle couple to the Standard Model.

Other tests can also be performed. For instance, neglecting direct CP-violation in the
decay amplitudes, one can write a "theory-independent" relation among
$\DDbar$ mixing amplitudes~\cite{Grossman:2009mn,Kagan:2009gb},
\beq
\frac{x}{y} = \frac{1-|q/p|}{\tan\phi} 
\eeq
Current experimental results $x/y \approx 0.8 \pm 0.3$ imply that amount of CP-violation in the 
$\DDbar$ mixing matrix is comparable to CP-violation in the interference of decays and mixing
amplitudes. An extensive study of exclusive decays should be 
performed~\cite{Grossman:2006jg}, which could also shed some light on how large
CP-violation in charm decay amplitudes could be. Finally, new observables, such as
CP-violating "untagged" decay asymmetries~\cite{Petrov:2004gs} should be studied in
hadronic decays~\cite{Ryd:2009uf} of charmed mesons.

\section{Conclusions}

With first results from the LHC experiments coming out this year, we are eagerly awaiting 
discoveries of new particles and interactions at the TeV scale. Their proper {\it identification} 
is an important task that will require inputs from collider, low-energy and astrophysical 
experiments. Constraints on indirect effects of New Physics at flavor factories will help to 
distinguish among models possibly observed at the LHC. I reviewed recent progress in 
theoretical understanding of NP constraints in charm transitions, which were chiefly
driven by recent experimental observation of $\DDbar$ mixing as well as experimental studies of
other charm meson transitions. With many LHC-favorite models already receiving interesting 
constraints from charm physics, new experimental results, especially in the studies 
of CP-violation, will be be indispensable for physics of the LHC era.

\section*{Acknowledgments}

This work was supported in part by the U.S.\ National Science Foundation
CAREER Award PHY--0547794, and by the U.S.\ Department of Energy under Contract
DE-FG02-96ER41005.


\end{document}